# All-Optical Nonzero-Field Vector Magnetic Sensor For Magnetoencephalography

M.V. Petrenko[1], A.S. Pazgalev[1], and A.K. Vershovskii[1]

[1]Ioffe Institute, Russian Academy of Sciences, St. Petersburg, 194021 Russia
e-mail address: antver@mail.ioffe.ru

We present the concept and the results of an investigation of an all-optical vector magnetic field sensor scheme developed for biological applications such as non-zero field magnetoencephalography and magnetocardiography. The scheme differs from the classical two-beam Bell-Bloom scheme in that the detecting laser beam is split into two beams, which are introduced into the cell in orthogonal directions, and the ratio of the amplitudes of the magnetic resonance signals in these beams and their phase difference are measured; strong optical pumping from the lower hyperfine level of the ground state ensures the resonance line narrowing, and detection in two beams is carried out in a balanced schemes by measuring the beam polarization rotation. The proposed sensor is compact, resistant to variations of parameters of laser radiation and highly sensitive to the angle of deflection of the magnetic field vector – with an estimated scalar sensitivity of the order of 16 fT/Hz$^{1/2}$ in 8×8×8 mm$^3$ cell, an angular sensitivity of $4 \cdot 10^{-7}$ rad, or 0.08", was demonstrated.

## INTRODUCTION

In recent decades, an avalanche growth of interest in the problems of studying ultra-weak magnetic fields generated by brain activity (magnetoencephalography, "MEG") has been observed. This is due to the advent of compact optical magnetic field (MF) sensors that can replace expensive superconducting SQUID systems [1] and overcome their inherent limitations. The principle of operation of these sensors is based on the effect of optically detected magnetic resonance (MR) in alkali atoms [2–7]; relaxation in the cells is suppressed by applying a special coating [8,9] or adding a buffer gas [10,11], with the latter option allowing the cell to be maximally compacted.

The highest sensitivity to date is demonstrated by zero-MF optical sensors using the SERF effect [2,12–15]. However, the growing need for sensors capable of operating outside stationary magnetically isolated rooms has given impetus to research on the possibility of adapting non-zero MF sensors to MEG tasks [7,16,8]. These sensors are initially characterized by somewhat worse sensitivity, but their use can significantly reduce the requirements for the parameters of an external MF [7].

Non-zero MF alkaline sensors atoms are characterized by one more drawback: they measure the modulus of the magnetic field, but not its components. In MEG, this feature leads to the fact that, in the presence of a relatively strong external MF $\mathbf{B_0}$, they respond only to the component of the brain field which is collinear to the $\mathbf{B_0}$ vector. Since the field generated by systems of magnetic shields and/or magnetic coils is spatially inhomogeneous, the task of interpreting the MEG signal in a non-zero MF requires knowing the direction of the vector $\mathbf{B_0}$ at the location of each sensor.

At the time of writing, there are a number of techniques that convert a scalar non-zero field sensor into a vector one, based both on measuring the response to a calibrated change in the transverse MF components [17–20], and on the observation of two or three components of the magnetic moment [21–23]. The latter technique was first proposed in [24], and then in our earlier work [25] a modification that makes it possible to exclude the system of magnetic coils from the scheme was described. In this work, we have used the principles outlined in [24,25], modifying them for use in a compact all-optical compact sensor.

In this paper, we present a scheme of an all-optical non-zero field vector magnetometer-variometer based on a scalar non-zero field sensor built according to the Bell-Bloom scheme [16,26–28]. The scheme differs from the classical two-beam Bell-Bloom scheme primarily in that the detecting laser beam is split into two beams, which propagate through the cell in orthogonal directions. The ratio of the amplitudes of the MR signals in these beams and the difference in their phases are measured. The proposed sensor is characterized by compactness (since an additional beam can be added to the scheme without increasing its size), high sensitivity to the MF vector deviation at the level of 0.1", and resistance to the laser radiation parameters' variations.

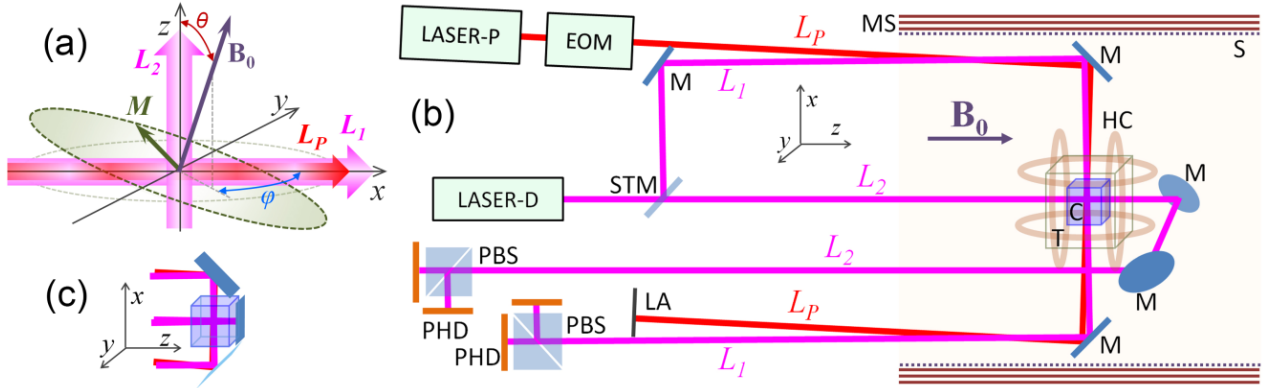

Fig. 1. (a) Pump and detection scheme: $L_p$ is the pumping beam, $L_1$, $L_2$ are the detecting beams, $\mathbf{B_0}$ is the external magnetic field vector, $\mathbf{M}$ is the magnetic moment. (b) Simplified scheme of the experiment: the sensor includes LASER-P – pumping light source, LASER-D – detecting light source, EOM – electro-optical modulator, M – mirror, STM – semitransparent mirror, C – gas cell with Cs vapor, T – thermostat, LA – light absorber, PBS – polarizing beamsplitter cube, PHD – balanced photodetectors. Elements of the experimental setup (not included in the sensor scheme): MS – magnetic shield, S – solenoid, HC – two-coordinate Helmholtz coils system. (c) Possible modification of the scheme: a return mirror in the beam $L_2$.

## BASIC PRINCIPLES

The scheme under consideration uses a number of principles previously used by us in [28].

First, for pumping and excitation of the MR, it uses a beam with ellipticity modulated at a frequency ($\omega \approx \omega_0$, where $\omega_0$ is the Larmor precession frequency). Beam polarization changes from left to right circular. Secondly, we use combined hyperfine-Zeeman pumping, which was first proposed in [29] and received a theoretical basis in [30]. The pump beam frequency is tuned to resonance with the $D_1$ transitions of the alkali metal line, which couple the hyperfine level $F = I-½$ (where $I$ is the nuclear spin) of $S_{1/2}$ state with the levels $F' = I\pm½$ of $P_{1/2}$ state [29,30]. Zeeman pumping of $F = I+½$, $m_F = F$ level is due to i) partial overlapping of the optical absorption contours of two ground-state hyperfine transitions, and ii) conservation of the nuclear component of the moment in the excited state [30]. Thirdly, we use strong optical pumping, which allows us to collect most of the atoms at the level $F = I + ½$, $m_F = F$. W. Happer called this state "end-state" or "stretched" and showed [31] that the spin-exchange rate in this state can significantly decrease.

Let $L_P$ denote the pumping beam (Fig. 1(a)). It is directed along the $x$ axis and modulated at the frequency $\omega$. The magnetic field $\mathbf{B_0}$ is initially directed along the $z$ axis. Pumping creates a magnetic moment $\mathbf{M}$ along beam $L_P$, whose component perpendicular to $\mathbf{B_0}$ precesses around $\mathbf{B_0}$. The pumping efficiency in the first approximation is proportional to the sine of the angle between $L_P$ and $\mathbf{B_0}$. $L_1$, $L_2$ are the detecting beams, generated by the same laser. Their intensities are $I_1$ and $I_2$, respectively. The beams are linearly polarized and far detuned from the $D_1$ absorption line. Beam $L_1$ is directed along the $x$ axis, beam $L_2$ is directed along the $z$ axis.

The signals of the rotation of the polarization azimuth of the detecting beams $L_1$ and $L_2$ are detected at the frequency $\omega$. The method of synchronous detection determines their amplitudes $S_1$, $S_2$ and phases $\psi_1$, $\psi_2$ with respect to the modulation signal. The amplitudes are proportional to the projections of the rotating component of $\mathbf{M}$ on the direction of their propagation (the so-called $M_X$ scheme [32,33]). The $S_1$ signal, as in a conventional $M_X$ magnetometer, is used to tune the frequency $\omega$ to the MR frequency $\omega_0$. In addition, it is used as a reference signal in the processing of the $S_2$ signal, which makes it possible to determine the MF deflection angles regardless of the $L_P$, $L_1$, $L_2$ parameters.

Let's switch to a spherical coordinate system defined by the polar angle $\theta$ and the azimuthal angle $\varphi$ (Fig. 1a). It is easy to show that, within this simplified model, the angles $\theta$ and $\varphi$ are expressed in terms of the phase difference $\psi_1$ and $\psi_2$ and the ratio $S$ of the signals $S_1$ and $S_2$, normalized to the ratio of the intensities of the corresponding beams:

$$\varphi = \psi_2 - \psi_1$$
$$S \equiv \frac{|S_2|}{|S_1|} \frac{I_1}{I_2} = k \frac{\sin\theta}{\sqrt{1-\sin^2(\theta)\cos^2(\varphi)}}, \quad (1)$$

where $k$ is the scale factor that takes into account the inequality of the volumes of intersection of beams $L_1$ and $L_2$ with the pumping beam $L_P$.

For $\varphi = 0°$ $S = \text{tg}(\theta)$, for $\varphi = 90°$ $S = \sin(\theta)$; at small $\theta$ $S \approx \theta$ for any $\varphi$.

Let us estimate the achievable sensitivities to variations in the transverse field components. In the ideal case ($\theta \approx 0$, $I_1 \approx I_2$), neglecting the second-order terms, we obtain:

$$\delta B_x = \delta B_y \approx \frac{1}{k}\frac{\omega}{\Gamma}\delta B, \quad (2)$$

where $\Gamma$ is the relaxation rate (MR half-width), and the scalar sensitivity $\delta B \approx \delta B_z$ is determined by the well-known expression

$$\delta B = \frac{\Gamma}{\gamma} \frac{n}{S_1} \quad (3)$$

where $\gamma$ is the gyromagnetic ratio and $n$ is the spectral density of the principal quantum noise (atomic projection noise plus photon shot noise; in optimized schemes, these two noises are approximately equal in spectral power). The transverse sensitivity $\delta B_\perp = \delta B_x = \delta B_y$ turns out to be $(1/k)(\omega/\Gamma)$ times worse than $\delta B$.

Let us estimate the angular sensitivity for small $\theta$:

$$\delta\theta \approx \frac{1}{k}\left(\frac{S_1}{n}\right)^{-1}, \quad (4)$$

$$\delta\phi \approx \left(\frac{S_2}{n}\right)^{-1} \approx \frac{\delta\theta}{\theta}. \quad (5)$$

As expected, the polar angle sensitivity is independent of $B$, and is determined by the signal-to-noise ratio in the $L_1$ beam. Uncertainty $\delta\varphi$ at large $\theta$ approaches $\delta\theta$, but at small $\theta$ tends to infinity; this is due not to the features of the proposed scheme, but to the general properties of the spherical coordinate system.

## EXPERIMENTAL SETUP

Measurements were made on the setup [28,34], modified in accordance with the task of the experiment (Fig. 2(b)). The light sources were diode lasers with an external cavity manufactured by VitaWawe. The pumping laser frequency was tuned to the Cs $D_1$ transitions, which couple the hyperfine level $F = 3$ of the $S_{1/2}$ state to the levels $F' = 3,4$ of the $P_{1/2}$ state. The frequency of the detecting laser was tuned 15–20 GHz down from the $D_1$ line transitions linking the level $F = 4$ of the $S_{1/2}$ state with the levels $F' = 3,4$ of the $P_{1/2}$ state.

A Thorlabs electro-optical modulator was used to modulate the pumping light polarization. The available pumping light power at the cell input was 5.4 mW, the detecting light power ($L_1+L_2$) was set at 4.9 mW. The sensitive element of the sensor was a cubic cell $8 \times 8 \times 8$ mm$^3$ containing saturated cesium vapor and nitrogen at a pressure of ~100 Torr. A thermostat with cells and a heater were placed in the central region of a multilayer magnetic shield, in which the MF induction varied within 0.1–12 µT.

We used a configuration with all three beams passing through the cell once, which required the installation of two mirrors behind the cell at an angle of 45° to beam $L_2$ (Fig. 2(b)), and resulted in an increase in the volume of the sensor. However, the compactness of the sensor can be preserved (and, at the same time, its vector sensitivity can be increased) by installing a mirror (Fig.1(c)) in the thermostat that returns the $L_2$ beam back through the entrance window at a small angle (as was done, for example, in [35]). In this case, the length of interaction of the beam $L_2$ with the optically oriented media will double. This will lead to a proportional increase in the signal $S_2$ (and a corresponding change in $k$), but, possibly, will require an increase in the detuning of the detecting laser from the optical transition. The same technique can be applied to the beams $L_P$ and $L_1$.

The noise spectrum of the vector sensor was studied by applying a calibrated transverse field with an amplitude of 100 pT rms, oscillating at 10 Hz.

## RESULTS

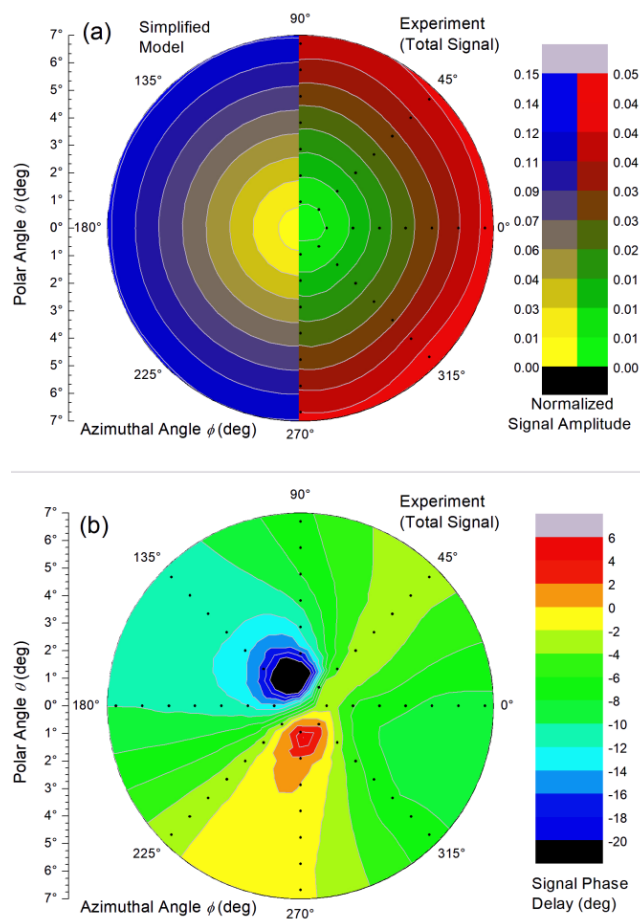

Fig. 2. (a) Amplitude $S$ of the normalized vector signal: on the left – predictions of the simplified model (Eq. (1)), on the right – experiment. (b) Phase delay $\psi_2 - \psi_1 - \varphi$ of the normalized vector signal. $\Gamma = 100$ nT, $B_0 = 3000$ nT.

A typical distribution of the parameters of the vector signal $S$ at small polar angles $\theta$ and magnetic fields, significantly exceeding the half-width $\Gamma$ of the MR, is shown in Fig. 2. It can be seen that the signal amplitude is fairly well described by Eq. (1), and the value of $k$ is $k = 0.40 \pm 0.01$; in accordance with Eq. (5), the measurement uncertainty $\delta\varphi$ increases with decreasing $\theta$.

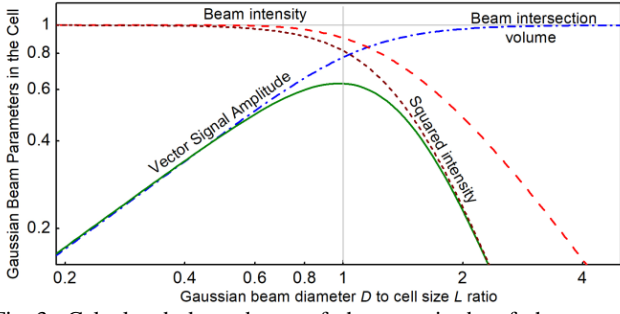

Fig. 3. Calculated dependence of the magnitude of the vector signal and the intensity of the Gaussian beam of diameter $D$ in a cubic cell with face length $L$.

By numerical integration, we calculated the overlap integral of the beams $L_P$ and $L_2$ under the assumption that they are characterized by the same diameter $D$ (Fig. 3), and approximated the resulting dependence by the expression

$$k = \frac{r/1.16}{\left(1+(r/1.16)^4\right)^{\frac{1}{4}}}, \quad (6)$$

where $r = D/L$ is the ratio of the diameter of the beams to the length of the edge of the cubic cell $L$. An increase in $D$ leads to an increase in the vector sensitivity, but also to light losses: for each beam, the dependence of the intensity of the light entering the cell on $r$ is described by the approximate formula

$$I = \frac{1}{\left(1+(r/1.4)^4\right)^{0.44}}, \quad (7)$$

the signal being proportional to the product of the intensities of the pump and detection beams. The final dependence of the magnitude of the vector signal on $r$ has a maximum at $r \approx 1$. These estimates do not take into account changes in the scalar sensitivity, which, other parameters being constant, is proportional to the root of the number of interrogated atoms [36] and depends in a complex way on the pump intensity [28].

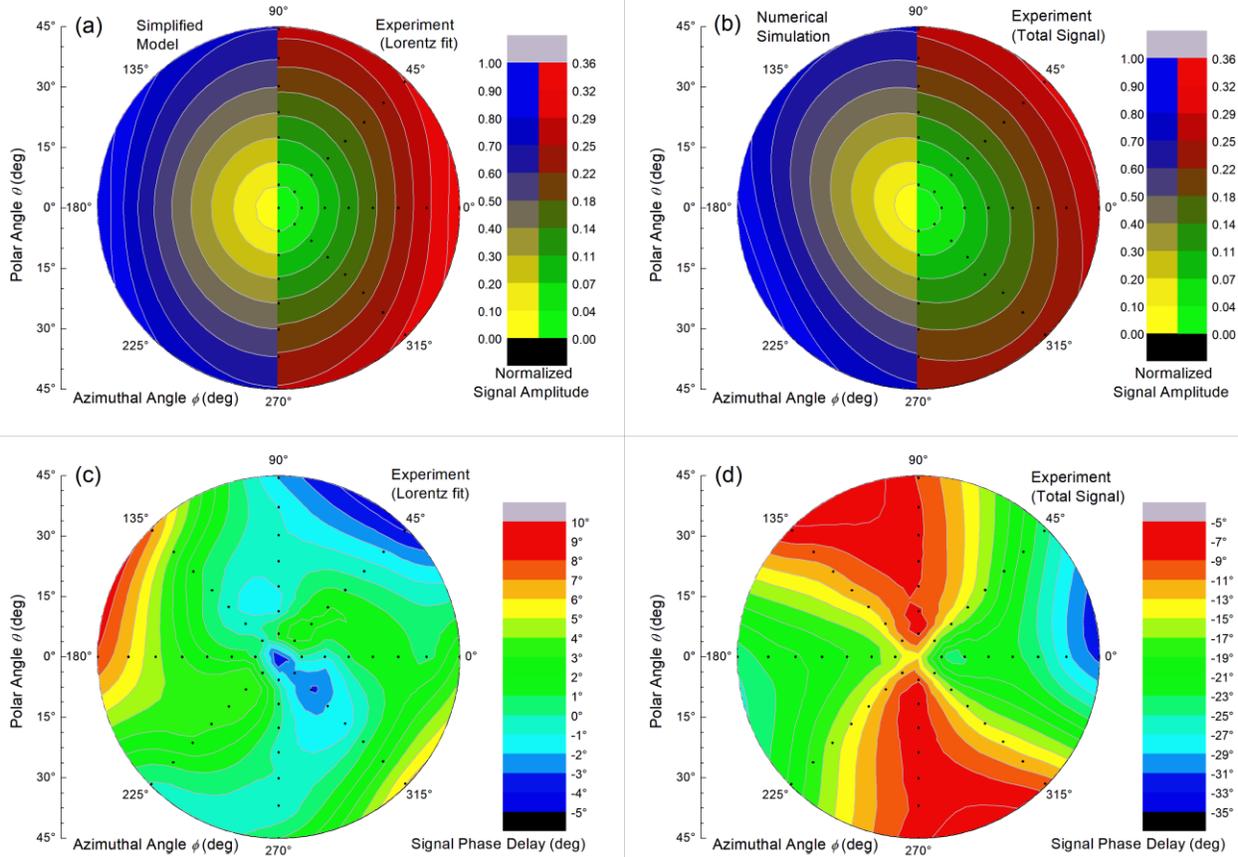

Fig. 4. (a) Amplitude $S_L$ of the Lorentzian part of the normalized vector signal: on the left, the predictions of the simplified model (Eq. (1)), on the right, the experiment. (b) Amplitude $S_T$ of the total normalized vector signal: on the left, the result of numerical simulation, on the right, the experiment. (c) Phase delay $\psi_{2L} - \psi_{1L} - \varphi$ of the part of the Lorentzian part of the normalized vector signal. (d) Phase delay $\psi_{2T} - \psi_{1T} - \varphi$ of the total normalized vector signal. $\Gamma = 100$ nT, $B_0 = 500$ nT.

In our experiment, $r = 0.575$, and $k = 0.40 \pm 0.01$, which is quite close to the value of $k(r) = 0.49$ predicted by Eq. (6). The remaining difference can be explained by the depolarizing properties of the mirrors located on the path of the beam after the cell: the phase delay introduced by the mirrors leads to a partial transformation of the rotation of the polarization azimuth into its ellipticity.

The difference between the experiment and the predictions of the simplified theory is manifested at relatively small fields and large deflection angles $\theta$ (Fig. 4, $B_0 \approx 5\Gamma/\gamma$).

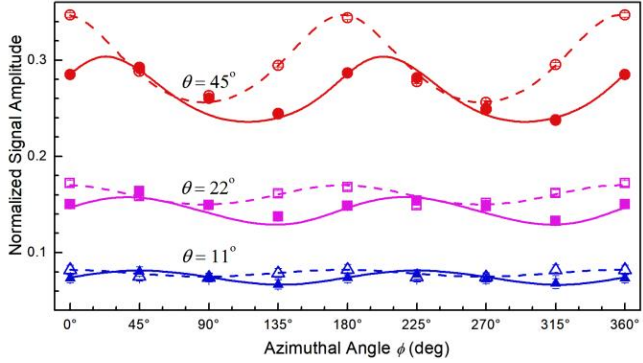

Fig. 5. Dependence of the vector signals $S_L$, $S_T$ at $\Gamma = 100$ nT, $B_0 = 500$ nT, $\theta = 11°, 22°, 45°$ on the azimuthal angle $\varphi$: dashed lines – predictions of the simplified model (Eq. (1)), open symbols – experiment (the amplitude $S_L$ of the Lorentzian contour); solid lines are the result of numerical simulation, filled symbols are the experiment (the amplitude $S_T$ of the total signal).

The signals measured in our experiment when scanning the frequency $\omega$ in the vicinity of $\omega_0$ were approximated by Lorentzian contours with quadratic baselines. The amplitudes $S_{2L}$, $S_{1L}$ (and their ratio $S_L$) of the contours and phases $\psi_{2L}$, $\psi_{1L}$ were calculated. The amplitudes $S_{2T}$, $S_{1T}$ (and their ratio $S_T$) and the phases $\psi_{2T}$, $\psi_{1T}$ of the total signals (i.e., the signals that are measured in a real sensor) were also calculated.

The simplified model (Eq. (1)) describes well the parameters of the Lorentzian contour (Fig. 4(a)), but does not provide a correct description of the total signal $S_T$. To do this, we carried out a numerical simulation: the dynamics of the magnetic moment was described by a system of non-stationary Bloch equations under conditions of periodic pumping. As can be seen from Fig. 4(b) and Fig. 5(b), the numerical simulation adequately describes the total signal $S_T$.

The calculated and experimental spectra of the magnetic resonance signal in a weak field are shown in Fig. 6. As it turned out, the distortion of the shape of the resonance is due to another signal, which is maximum at zero modulation frequency. It arises when the vector $\mathbf{B_0}$ deviates in the $Oxy$ plane, and is due to the non-precessing component of the magnetic moment, oriented along $\mathbf{B_0}$ and changing its sign twice during the modulation period. The wing of this signal is added to the Lorentzian contour of the $M_X$ resonance, which leads to its distortion and phase shift.

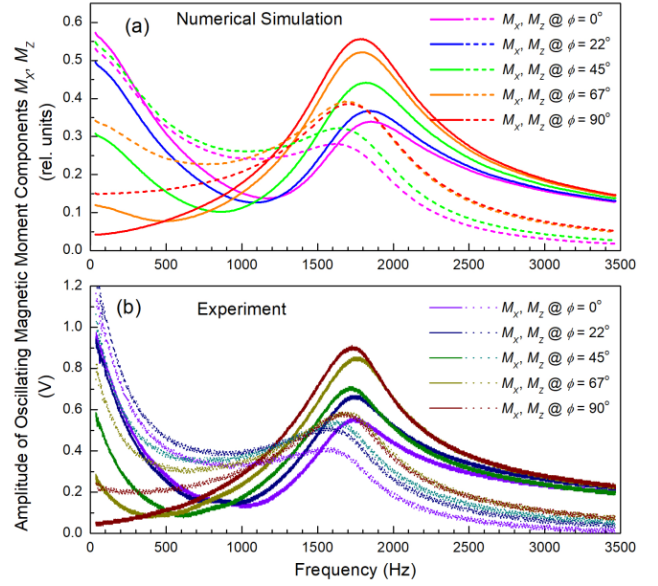

Fig. 6. Dependence of the $x$- and $z$-components of the magnetic moment $\mathbf{M}$ on the pump beam modulation frequency: (a) the result of numerical simulation, (b) the experiment (the measured values of $M_Z$ are multiplied by $1/k = 2.5$ to take into account the difference in the sensitivities of the $Z$ and $X$ channels). $\Gamma = 100$ nT, $B_0 = 500$ nT, $\theta = 45°$.

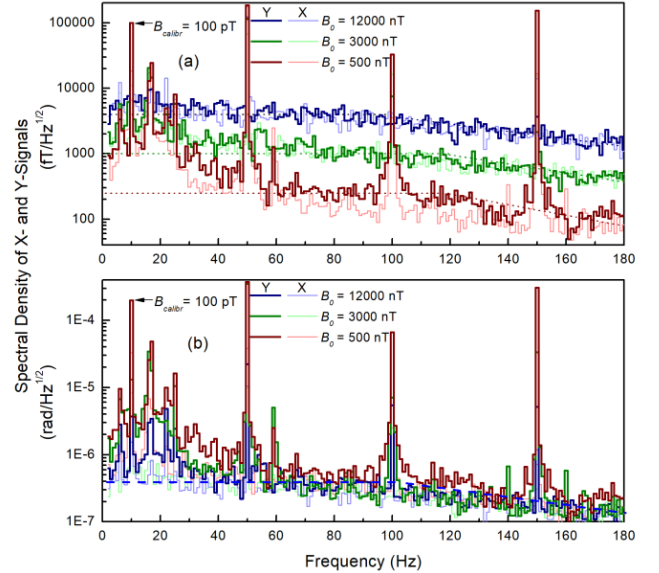

Fig. 7. Spectral dependence of vector signal noise ($X$ and $Y$ channels) at three values of the field $B_0$: (a) in magnetic field units, (b) in angular units. A calibrated magnetic field disturbance $B_Y = 100$ pT rms is applied to coil $Y$. 50 Hz signals and their harmonics are mains interference.

Fig.7 shows the results of measurements of the transverse noise spectral density in a magnetic shield.

## DISCUSSION

From the above results, it follows that the complex nature of the spectrum of the signals $S_{1T}$ and $S_{2T}$ does not lead to a deterioration in the parameters of the vector sensor, but requires corrections for the calculation of $\theta$ and $\varphi$.

Vector sensitivity is measured at the level of (0.39±0.1) μrad (Fig.7), or 0.08″. As follows from Eqs. (2), (4), (5), the sensitivity to the transverse field components $\delta B_\perp$ deteriorates with increasing $B_0$, while the angular sensitivity does not depend on $B_0$.

Table I. *The results of the study of the parameters of the vector sensor.*

| # | Parameter | Designation | Expected Value | Experiment | Units |
|---|---|---|---|---|---|
| 1 | Cubic cell edge length | $L$ | - | 8.0±0.2 | mm |
| 2 | Pump-detection beam diameter | $D$ | - | 4.6±0.3 | mm |
| 3 | The power of the beams $L_1$, $L_2$ on the photodetectors | $I_1$, $I_2$ | - | 0.70±0.05 | mW |
| 4 | Amplitude of the signal in $L_1$ at $B_0$ = 500 nT, $\theta$ = 0 | $S_1$ | - | 0.19±0.01 | mA |
| 5 | Magnetic resonance half-width | $\Gamma$ | - | 2π·(350±6) | Hz |
| 6 | Shot noise spectral density | $n_{sh}$ | 12.6 | - | pA/Hz$^{1/2}$ |
| 7 | Signal-to-noise ratio in $L_1$ (shot-noise) | $S_1/n_{sh}$ | 1.5·10$^7$ | - | |
| 8 | Ultimate sensitivity (shot) | $\delta B_{sh}$ | 6 | - | fT/Hz$^{1/2}$ |
| 9 | Total noise spectral density | $n$ | 3 $n_{sh}$ | - | |
| 10 | Ultimate sensitivity (total noise) | $\delta B$ | 18 | 16±3 * | fT/Hz$^{1/2}$ |
| 11 | Signal-to-noise ratio in L1 (total noise) | $S_1/n$ | 5·10$^6$ | 6.2±1.3·10$^6$ * | |
| 12 | Scale factor | $k$ | 0.49 | 0.40±0.01 | |
| 13 | Angular Resolution | $\delta\theta$ | 0.41 | 0.39±0.08 ** | μrad |
| 14 | Sensitivity to field components at $B_0$ = 500 nT | $\delta B_\perp$ | 230 | 220±50 ** | fT/Hz$^{1/2}$ |
| 15 | MF transverse noise spectral density | $n_{MF}$ | - | 130±50 ** | fT/Hz$^{1/2}$ |
| 16 | Angular Resolution (corrected) | $\Delta\theta'$ | 0.41 | 0.35±0.08 *** | μrad |
| 17 | Sensitivity to field components at $B_0$ = 500 nT (corr.) | $\delta B_\perp'$ | 230 | 180±80 *** | fT/Hz$^{1/2}$ |

\* Derived from the value $\delta\theta$ given in line 13 using Eqs. (4), (3).
\*\* Derived from the experimental data (Fig. 7) taken in the frequency range of 65–90 Hz.
\*\*\* Derived from the values $\delta\theta$ and $\delta B_\perp$ (lines 13-14): the contribution of the MF transverse noise $n_{MF}$ (line 15) is excluded.

The results of the study of the parameters of the vector sensor are shown in Table I. The estimated values given in the table are calculated using equations (2)–(6) and the values in lines 1–5 of Table I; it is also taken into account that, according to Refs. [28,34], the magnitude of real noise (including atomic projection noise and technical noise of laser radiation) under our experimental conditions is approximately three times higher than the shot noise level.

Since the typical value of the noise of the longitudinal field component in the frequency range free from acoustic disturbances in our shield is ~0.6 pT/Hz$^{1/2}$ [37] (which is about five times the noise of the transverse field components), a direct measurement of the real signal-to-noise ratio and the real scalar sensitivity of the sensor is possible only in the gradiometric scheme [37,38].

However, the use of the vector scheme provides us with an indirect tool for measuring these quantities. From the results of measuring the angular resolution $\delta\theta$ in a strong enough MF, using Eq. (4), one can determine the real signal-to-noise ratio $S_1/n$, and then, using Eq. (3), the real scalar sensitivity $\delta B$. The estimated values given in Table I (lines 10–14) are in excellent agreement with the experimental results.

According to Eqs. (2)–(5), the achieved sensitivity to the field components can be improved more than twofold by doubling the beam diameters in the cell with a corresponding fourfold increase in the integrated power. It should also be taken into account that in this case, the scalar sensitivity will double [36] (proportionally to the root of the number of interrogated atoms), and the overall improvement will be almost fivefold.

This expected vector sensitivity approaches that required of the three-component sensor for operation in MEG complexes; this is indirectly confirmed by the results of [39]. In any case, the achieved angular sensitivity is certainly sufficient to measure the direction of the external MF vector in non-zero field MEG systems.

Sensitivity $\delta B_\perp$ can be further improved by reducing the field to $B_0 \approx \Gamma/\gamma$; in our experiment, this is prevented by low-frequency noise and modulations of laser radiation. In this case, the transverse sensitivity can reach its limit value of $\delta B_\perp \approx \delta B$.

## CONCLUSIONS

We have demonstrated the operability of the proposed concept, and have shown that the parameters of the scheme correspond to the calculated ones. Due to the use of a reference detecting beam, the measurement result of the magnetic field deflection angles is not affected by variations in the spectral parameters of the pumping and detecting beams. The proposed scheme can be implemented in the same sensor dimensions as the Bell-Bloom scheme or the standard two-beam $M_X$ magnetometer scheme. The demonstrated parameters allow real-time measurement of the distribution of external MF vector directions in non-zero field MEG complexes. A further (almost fivefold) improvement in angular sensitivity can be achieved by increasing the beam diameters with a corresponding increase in the pump and detection powers. This will require the use of high-power, low-noise pump and detection laser sources; but in this case, the sensitivity of the proposed sensor will be sufficient to measure all three components of the MEG vector signal in a non-zero magnetic field.

An additional, but useful in the laboratory conditions. result of this work is a new method for determining the scalar sensitivity of the sensor under conditions when the MF noise significantly exceeds the level of intrinsic noise.